\documentstyle[12pt]{article}
\newcommand{\be}{\begin{equation}}
\newcommand{\ee}{\end{equation}}
\newcommand{\ba}{\begin{eqnarray}}
\newcommand{\ea}{\end{eqnarray}}

\begin{document}
\begin{center}
 {\bf\Large{
   Killing-Yano tensors and surface terms }}
\end{center}
\begin{center}
{\bf Dumitru Baleanu}\footnote[1]{ On leave of absence from
Institute of Space Sciences, P.O BOX, MG-23, R 76900
Magurele-Bucharest, Romania, E-mail: dumitru@cankaya.edu.tr}, {\bf
{\"O}zlem Defterli\footnote [2]{E-Mail:~~defterli@cankaya.edu.tr}}
\end{center}
\begin{center}
Department of Mathematics and Computer Sciences, Faculty of Arts
and Sciences, Cankaya University-06530, Ankara , Turkey
\end{center}
\begin{abstract}
New geometries were obtained by adding a suitable surface term
involving the components of the angular momentum to the
corresponding free Lagrangians.Killing vectors, Killing-Yano  and
Killing tensors of the obtained manifolds were investigated.
 \end{abstract}

\section{Introduction}

 Killing-Yano (KY) tensors introduced by Yano \cite{yano52} play an important role in the Dirac theory on the curved
spacetimes \cite{carter}.
 A new supersymmetry corresponding to (KY) tensor was
found in black-hole solutions of the Kerr-Newman type \cite{gib}.
 (KY) tensors are crucial in supersymmetric extension of charged point
particle's motion in investigation of  the symmetries of
gravitational and electromagnetic fields \cite{tanimoto}. (KY)
tensors of order three are special cases of Lax tensors introduced
by Rosquist \cite{ros97}. A (KY) tensor of order two generates a
Killing tensor and, in some cases, it produces new constants of
motion on a curved manifold \cite{coll,brian, book}. In the last
yers a huge effort was devoted for analyzing the importance of
(KY) tensors \cite{hol95,hol96,balnambu,holt99, jacek, holten2000,
visinescu2003} in several areas but there are relatively few
manifolds of physical interest admitting these tensors. This
drawback is mainly because (KY) tensors are antisymmetric and
their equations impose restrictions on the manifold structure
\cite{book}. Despite of this fact there were  some attempts to
find new geometries admitting (KY) tensors of order two or three.
For example, the three- particle open Toda lattice was geometrized
by a suitable canonical transformation and it was realized as the
geodesic system of a certain Riemannian geometry \cite{ros98}.
Adding a time-like dimension a four-dimensional space-times
admitting  two Killing vector fields were found \cite{ros98}.

Motivated by the above results we decided to construct integrable
geometries admitting (KY) and Killing tensors by adding a surface
term \cite{vergara}, \cite{montesino} to a known free Lagrangian.

The main aim of this paper is to add, using suitable Lagrangian
multipliers, the components of the angular momenta to a free
Lagrangian in two or three dimensions and to study the hidden
symmetries of the induced manifolds.

 The plan of the paper is as follows:

 In Sec.2 generic and non-generic symmetries of the extended Lagrangians are
 investigated.
 Sec.3 deals with symmetries of new geometries induced by the motion on a
 sphere.
 Our conclusions are given in Sec. 4.

\section{Extended Lagrangians and their\\
 corresponding geometries}

Let us assume that a given free Lagrangian $L({\dot q^i},q^i)$
admits a set of constants of motion denoted by $L_{i}, i=1,\cdots
3$. If we  add the components of the angular momentum
corresponding to L, the extended Lagrangian $L^{'}=
L+{\dot\lambda^i}L_i, i=1,\cdots 3$ can be rewritten as
 $L^{'}=\frac{1}{2}a_{ij}{\dot q^{i}}{\dot q^{j}}$. Since $a_{ij}$ is
 symmetric by construction,
 the issue is to find a way to construct induced manifolds. In
 other words we are looking to find whether $a_{ij}$ is singular
 or not.
 If the matrix $a_{ij}$ is singular $L^{'}$
corresponds to a singular system \cite{murr}. Assuming that
$a_{ij}$ is a singular nxn matrix of rank n-1  we  obtain
non-singular symmetric matrices of order (n-1)x(n-1), where n will
be 3,5 and 6. The final step is to consider the obtained matrices
as metrics on the extended space and to investigate their generic
and non-generic symmetries.

\subsection{The nonsingular case}

 {\bf 2.1.1.} As a starting point let us consider the following Lagrangian

\begin{equation}\label{priimaa}
 L^{'}=\frac{1}{2}(\dot{x}^2+\dot{y}^2)+\dot{\lambda_{3}}(x\dot{y}-y\dot{x})
\end{equation}
From (\ref{priimaa}) we obtain
 $L^{'}=\frac{1}{2}a_{ij}{\dot q^{i}}{\dot q^{j}}$, where $a_{ij}$ is
 given by
 \be\label{matrice}
 a_{ij}=\left( \begin{array}{ccc}
 1 & 0 &-y \\
 0 & 1& x \\
-y & x& 0 \end{array}\right). \ee The corresponding Killing vector
is  V=(y, -x, 0).\\
 A (KY) is an antisymmetric tensor define as
\be\label{ky} D_{\lambda}f_{\mu\nu} + D_{\mu}f_{\lambda\nu}=0. \ee

  Solving (\ref{ky}) corresponding to (\ref{matrice})
 we obtained the following (KY) tensor
\begin{equation}\label{oz}
   f_{13}=0, f_{23}=-Cx\sqrt{x^2+y^2},f_{13}=Cy\sqrt{x^2+y^2},
\end{equation}
where C is a constant.

If a (KY) tensor exists, then a Killing tensor of order is
generated as \be\label{formula}
K_{\mu\nu}=f_{\mu\lambda}f_{\nu}^{\lambda}. \ee
 Using (\ref{oz}) and (\ref{formula})  a
 Killing tensor is constructed as
\be\label{ktky} K_{ij}=\left(
\begin{array}{ccc}
  y^2 & -xy & -y(y^2+x^2) \\
  -xy & x^2 & x(x^2+y^2) \\
  -y(y^2+x^2) & x(x^2+y^2) & 0 \\
\end{array}%
\right) \ee

Another method to obtain a Killing tensor is to solve the
corresponding equations \be\label{ten} D_\mu k_{\nu\lambda}+D_\nu
k_{\lambda\mu}+D_\lambda k_{\mu\nu}=0,\ee where $k_{\mu\nu}$ is a
symmetric tensor. Solving (\ref{ten}) corresponding to
(\ref{matrice}) we obtain a class of solutions given by
\be\label{mkt}
\begin{array}{cc}
k_{11}={1\over 2}y^2(C_2z+C_3)+C_1, k_{12}=-{1\over
2}xy(C_2z+C_3)\cr k_{13}=-{y\over 4}[(x^2+y^2)(C_2\arctan({x\over
y})-4C_4)+4C_1],k_{22}={1\over 2}x^2(zC_2+C_3)+C_1\cr
k_{23}={{x\over 4}}[(x^2+y^2)(C_2\arctan({x\over
y})-4C_4)+4C_1],k_{33}=0.
\end{array}
\ee

We observed that if $C_2=C_1=0,C_3={1\over 2}$ we reobtain the
solution from (\ref{ktky}). Choosing the appropriate values of the
constants $C_i,i=1,\cdots 4$, we obtain a set of non-singular
Killing tensors. These Killing tensors can be considered as
manifolds and we have so called geometric duality (for more
details see  Refs. \cite{hol96, hinter}). If $C_2=0$ the dual
metrics have the following forms \be\label{dfirst}
\begin{array}{c}
 k_{11}^{-1}={x^2\over
(x^2+y^2)C_1},k_{12}^{-1}={xy\over (x^2+y^2)C_1},
k_{13}^{-1}={y\over
(x^2+y^2)[C_4(x^2+y^2)-C_1]},k_{22}^{-1}={x^2\over
(y^2+y^2)C_1},\\
k_{23}^{-1}={x\over (x^2+y^2)[C_4(x^2+y^2)-C_1]},
k_{33}^{-1}=-{1\over 2}{{(x^2+y^2)C_3+2C_1}\over
(x^2+y^2)(C_1-C_4(x^2+y^2))^2 },
\end{array}
\ee

and the scalar curvatures corresponding to (\ref{dfirst}) are \be
R={2C_1C_4[5C_4(x^2+y^2)+2C_1]\over [-C_1+C_4(x^2+y^2)]^2}. \ee

 {\bf 2.1.2} Let us add  two
components of the angular momentum at a free, three-dimensional
Lagrangian. The extended Lagrangian becomes
\begin{equation}\label{doii}
L^{'}=\frac{1}{2}(\dot{x^{2}}+\dot{y^{2}}+\dot{z^{2}})+\dot{\lambda_{1}}(y\dot{z}-z\dot{y})+\dot{\lambda_{2}}(z\dot{x}-x\dot{z})
\end{equation}
and (\ref{doii}) we identify
 $a_{ij}$ as the following non-singular matrix
\begin{equation}\label{ccin} a_{ij}=\left(%
\begin{array}{ccccc}
  1 & 0 & 0 & 0 & z \\
  0 & 1 & 0 & -z & 0 \\
  0 & 0 & 1 & y & -x \\
  0 & -z & y & 0 & 0 \\
  z & 0 & -x & 0 &0 \\
\end{array}\right).
\end{equation}
The metric (\ref{ccin}) admits three Killing vectors
\be\label{kvect}
V_1=(y,-x,0,0,0),V_2=(0,-z,y,0,0),V_3=(z,0,-x,0,0). \ee
 In this case (KY) tensors components are given by \be
\begin{array}{c}
f_{15}=-G xy,f_{14}=G(z^2+y^2),f_{24}=-Gxy,f_{34}=-Gxz,\cr
f_{25}=G(x^2+z^2), f_{35}={-G xzy\over x},f_{12}=Cz,f_{13}=-Cy,
\end{array}
 \ee others zero. Here C and G are constants.
The corresponding Killing tensor has the following form

\be
K=\left(%
\begin{array}{ccccc}
G(-2C+G)(z^2+y^2)&GDxy&GDzx&0&G^2r^2z \\
GDxy&-GD(x^2+z^2)&GDzy&-r^2zG^2&0 \\
GDzx&GDzy&-GD(y^2+x^2)&G^2r^2y& -G^2r^2x\\
0&-G^2zr^2&G^2yr^2&0&0\\
G^2zr^2&0&-G^2xr^2&0&0
\end{array}%
\right), \ee
 where  D =2C+G and $r^2=x^2+y^2+z^2$.

\subsection{The singular case}

{\bf 2.2.1.} The final step is to add  all angular momentum
components at the Lagrangian of the free particle in
three-dimensions. In this case $L^{'}$ is given by
\begin{equation}\label{treii}
L^{'}=\frac{1}{2}(\dot{x^{2}}+\dot{y^{2}}+\dot{z^{2}})+\dot{\lambda_{1}}(y\dot{z}-z\dot{y})+\dot{\lambda_{2}}(z\dot{x}-x\dot{z})+\dot{\lambda_{3}}(x\dot{y}-y\dot{x})
\end{equation}
In compact form (\ref{treii}) is written as
$L^{'}=\frac{1}{2}a_{ij}{\dot q^{i}}{\dot q^{j}}$, where $a_{ij}$
is singular having the form \be
 a_{ij}=\left(%
\begin{array}{cccccc}\label{saispe}
  1 & 0 & 0 & 0 & z &-y\\
  0 & 1 & 0 & -z & 0& x\\
  0 & 0 & 1 & y & -x& 0\\
  0 & -z & y & 0 & 0& 0\\
  z & 0 & -x & 0 &0 & 0\\
 -y & x & 0  & 0 & 0& 0\\
\end{array}%
\right). \ee

 Since the rank  of (\ref{saispe}) is 5 we obtained three non-singular symmetric matrices corresponding to three non-zero minors.
 The first one is given by (\ref{ccin}) and the other two are as
 follows

\be\label{cunu} b_{\mu\nu}^{(2)}=\left(%
\begin{array}{ccccc}
  1 & 0 & 0 & 0 & -y \\
  0 & 1 & 0 & -z & x \\
  0 & 0 & 1 & y & 0 \\
  0 & -z & y & 0 & 0 \\
  -y & x & 0 & 0 &0 \\
\end{array}%
\right) \ee and \be\label{cdoi}
 b_{\mu\nu}^{(3)}=\left(%
\begin{array}{ccccc}
  1 & 0 & 0 &z & -y \\
  0 & 1 & 0 & 0 & x \\
  0 & 0 & 1 & -x & 0 \\
  z & 0 & -x & 0 & 0 \\
  -y & x & 0 & 0 &0 \\
\end{array}
\right) \ee By direct calculations we observed that (\ref{cunu})
and (\ref{cdoi}) admit three Killing vectors given by
(\ref{kvect}) and a (KY) tensor having the following non-zero
components \be f_{12}=z,f_{13}=-y,f_{23}=x. \ee

 \section {The motion on a sphere and its induced\\ geometries}
 It was proved in \cite{nambum} that the motion on a sphere admits four constants of motion, the Hamiltonian and three components of the angular momentum.
 In the following using the surface term we will generate four -dimensional manifolds.
 In this case the Lagrangian is given by

\begin{eqnarray}\label{sfera}
  L^{'} &=&\frac{1}{2}(1+\frac{x^2}{u})\dot{x}^2+\frac{1}{2}(1+\frac{y^2}{u})\dot{y}^2+\frac{xy}{u}\dot{x}
\dot{y}-\frac{xy}{\sqrt{u}}\dot{\lambda_{1}}\dot{x}+(\frac{x^2}{\sqrt{u}}+\sqrt{u})\dot{\lambda_{2}}\dot{x}\\
&-&(\frac{y^2}{\sqrt{u}}+\sqrt{u})\dot{\lambda_{1}}\dot{y}+\frac{xy}{\sqrt{u}}\dot{\lambda_{2}}\dot{y}+
x\dot{\lambda_{3}}\dot{y}-y\dot{\lambda_{3}}\dot{x},
\end{eqnarray}
where $u=1-x^2-y^2$. From (\ref{sfera})we identify the singular
matrix $a_{ij}$ as

\be\label{sfera}
 a_{ij}=\left(%
\begin{array}{ccccc}
1+\frac{x^2}{u} &\frac{xy}{u}  & -\frac{xy}{\sqrt{u}} &\frac{x^2}{\sqrt{u}}+\sqrt{u}  & -y \\
\frac{xy}{u}&  1+\frac{y^2}{u}& -\frac{y^2}{\sqrt{u}}-\sqrt{u} &\frac{xy}{\sqrt{u}}  & x \\
-\frac{xy}{\sqrt{u}}&-\frac{y^2}{\sqrt{u}}-\sqrt{u} &0  &0  & 0 \\
  \frac{x^2}{\sqrt{u}}+\sqrt{u} &\frac{xy}{\sqrt{u}} & 0 &0  & 0 \\
  -y & x &0  &0 & 0 \\
\end{array}%
\right). \ee Because (\ref{sfera}) is a singular matrix of  rank 4
we identify three symmetric minors of order four. If we treat
these minors as a metric we observed that they are not conformaly
flat but their scalar curvatures  are zero.

The first metric is given by
 \be\label{hgh}
g_{\mu\nu}^{(1)}=\left(%
\begin{array}{cccc}
1+\frac{x^{2}}{u}&\frac{xy}{u}&\sqrt{u}+\frac{x^2}{\sqrt{u}}&-y\\
\frac{xy}{u}&1+\frac{y^2}{u}&\frac{xy}{\sqrt{u}}&x\\
\sqrt{u}+\frac{x^2}{\sqrt{u}}&\frac{xy}{\sqrt{u}} & 0&0   \\
-y&x& 0 &0  \\
\end{array}%
\right). \ee The Killing vectors of (\ref{hgh}) has the following
components \be\label{lkj}
\begin{array}{cc}
V_{1}=(y,-x,0,0),V_{2}=(\sqrt{1-x^2-y^2}+{x^2\over {1-x^2-y^2}},{xy\over {1-x^2-y^2}},0,0)\\
V_{3}=(-{xy\over {1-x^2-y^2}},-\sqrt{1-x^2-y^2}-{y^2\over
{1-x^2-y^2}},0,0)\end{array}.\ee The next step is to investigate
its (KY) tensors. Solving (\ref{ky}) we obtain the following set
of solutions:

 {\bf a}.One-solution is $f_{21}=
{C_1\over{\sqrt{1-x^2-y^2}}}$, others zero.\\
 {\bf b}.Two-by-two
solution has the form: $f_{31}=f_{42}=C$,\\
{\bf c}. Three by three solution is  $f_{21}=
{C_1\over{\sqrt{-1+x^2+y^2}}}$ and $f_{31}=f_{42}=C$, where C and
$C_1$ are constants.

From (\ref{sfera}) another two metrics can be identified as
 \be\label{fgf}
g_{\mu\nu}^{(2)}=\left(%
\begin{array}{cccc}
1+\frac{x^{2}}{u}&\frac{xy}{u}&-\frac{xy}{\sqrt{u}}&-y\\
\frac{xy}{u}&1+\frac{y^2}{u}&-\sqrt{u}-\frac{y^2}{\sqrt{u}}&x\\
-\frac {xy}{\sqrt{u}}&-\sqrt{u}-\frac {y^2}{\sqrt{u}} & 0&0   \\
-y&x& 0 &0   \\
\end{array}%
\right) \ee

and

\be\label{nouaa} g_{\mu\rho}^{(3)}=\left(%
\begin{array}{cccc}
1+\frac{x^2}{u} &\frac{xy}{u}  & -\frac{xy}{\sqrt{u}} &\frac{x^2}{\sqrt{u}}+\sqrt{u} \\
\frac{xy}{u}&  1+\frac{y^2}{u}& -\frac{y^2}{\sqrt{u}}-\sqrt{u} &\frac{xy}{\sqrt{u}} \\
-\frac{xy}{\sqrt{u}}&-\frac{y^2}{\sqrt{u}}-\sqrt{u} &0  &0   \\
  \frac{x^2}{\sqrt{u}}+\sqrt{u} &\frac{xy}{\sqrt{u}} & 0 &0  \\
\end{array}%
\right) \ee respectively. By direct calculations we obtained that
(\ref{fgf}) and (\ref{nouaa}) admit the same Killing vector as in
(\ref{lkj}). Solving (\ref{ky}) corresponding to (\ref{fgf}) and
(\ref{nouaa}) we find
  one non-zero component of (KY) tensor as\\
  \be
f_{21}={C_1\over{\sqrt{1-x^2-y^2}}}. \ee

\section{Conclusions}

 Integrable geometries were reported by adding a surface term involving the components of the angular
momentum to a given free Lagrangian. The existence of Killing
vectors, (KY) and Killing tensors is investigated and in all cases
a solution is presented.

The first step was to add ,to a free two-dimensional Lagrangian, a
surface term involving  the third component of the angular
momentum. In this case a three -dimensional  metric was induced.
This metric is conformaly flat but its duals are not.

Increasing the number of dimensions to three and adding a surface
term involving  two components of the angular momentum we obtained
geometries, in four and five dimensions. The obtained induced
manifolds are not conformaly flat but all of them have Ricci
scalar zero.

If we add a surface term involving all components of the angular
momentum to a three dimensional free Lagrangian we observed that a
singular  matrix $a_{ij}$ arises. We identify three symmetric
minors of this metric and we investigated the existence of Killing
vectors , (KY) and Killing tensors corresponding to those induced
manifolds. We observed that the obtained manifolds admit the same
Killing vectors but  different  (KY) solutions.

Finally, the geometries induced by the motion on a sphere are
investigated and a four dimensional induced manifolds were
obtained. As in the previous case the manifolds admit the same
Killing vectors but different (KY) tensors.

 \section {Acknowledgments}
One of the authors (D. B.) would like to thank M. Henneaux, B.
Edgar and M. Montesinos for their helpful discussions.
 This work is partially supported by
the Scientific and Technical Research Council of Turkey.

\end{document}